\documentclass[prl,twocolumn,showpacs,psfig]{revtex4}
\usepackage{amsfonts,amsmath,amssymb}
\usepackage{graphicx}

\begin{document}

\newlength{\figwidth}
\setlength{\figwidth}{0.45\textwidth}

\preprint{\noindent Prepare for Physical Review Letter, version
0.15,
\hfill \today} \ \\ \ \\

\title{Probing Molecular Dynamics at Attosecond Resolution with Femtosecond Laser Pulses}

\author{X. M. Tong,\footnote{Contact: {xmtong@phys.ksu.edu}}
\ Z. X. Zhao, and C. D. Lin}

\affiliation{Physics Department, Kansas State University,
Manhattan, KS 66506-2601}

\begin{abstract}
The kinetic energy distribution of D$^+$ ions resulting from the
interaction of a femtosecond laser pulse with D$_2$ molecules is
calculated  based on the rescattering model. From analyzing the
molecular dynamics, it is shown that the recollision time between
the ionized electron and the D$_2^+$ ion can be read from the
D$^+$ kinetic energy peaks  to attosecond accuracy. We further
suggest that more precise reading of the clock can be achieved by
using shorter fs laser pulses (about 15fs).
\end{abstract}

\maketitle

Human experience shows that new areas of science and technology
open up with the ability to make measurements at increasingly
shorter time regime. With the advent of femtosecond (fs) lasers,
femtochemistry became possible where  chemical reaction dynamics
can be probed at the atomic scale \cite{Zewail00}. Clearly, fs
lasers cannot be used directly to probe electron dynamics which is
in the attosecond (as) regime. While a substantial effort is being
dedicated to developing single attosecond pulses
\cite{Drescher01,Hentschel01,Drescher02,Baltuska03}, presently few
laboratories have such lasers available.

An ingenious suggestion for performing measurements at attosecond
resolution with fs lasers was proposed by Corkum and his group
recently. Their results were reported in two recent publications,
here to be called  I \cite{Niikura02} and II \cite{Niikura03},
respectively. In their experiment, a 40 fs pulse, with mean
wavelength ranging from 800 nm to 1850 nm, and peak intensity of
about $1.5\times 10^{14}$ W/cm$^2$, was used to ionize a D$_2$
molecule to produce D$^+$ ion. It was assumed that D$_2$ was first
ionized near the   peak of the laser pulse to create a correlated
electronic and nuclear wave packet. Within a single optical cycle,
the electron was driven back to collide with D$_2^+$ and to excite
it to the excited $\sigma_u$ electronic state which subsequently
dissociated to produce D$^+$. The kinetic energy of the D$^+$ ion
reflects the internuclear distance, as well as the time when the
rescattering occurs. With proper laser intensity, both the initial
ionization and the rescattering are found to occur at time
interval of far less than one optical cycle, thus providing
attosecond temporal resolution, irrespective of the femtosecond
pulse duration of the laser. In II, by changing the wavelength of
the fs laser, they concluded that the dissociation dynamics of
D$_2^+$ can be used as a molecular clock and the clock can be read
with attosecond resolution.

To read the molecular clock accurately, the rescattering mechanism
which leads to the measurable D$^+$ kinetic energy distribution
has to be understood in details. In this Letter we report the main
conclusion of our careful analysis of the rescattering mechanism.
In contradiction to I and II, our analysis shows that the D$^+$
ions are not produced by the dissociation of the excited D$_2^+$
ions. We found that the excited D$_2^+$ ions are readily further
ionized by the laser and thus the D$^+$ ions measured in I and II
are mostly coming from the Coulomb explosion along the D$^+$ +
D$^+$ potential curve.  Our conclusion is consistent with the
recent experiment of Alnaser {\it et al.} \cite{Alnaser03} where
the branching ratio of D$^+$ produced from Coulomb explosion vs.
from dissociation was determined. Our analysis also shows that the
dominant peak of the D$^+$ kinetic energy distribution is from the
third return of the rescattering process, rather than from the
first return, as assumed in I and II.  This has the effect that
the previous molecular clock was not read correctly. From our
analysis we further suggest that a more precise reading of the
clock can be achieved with a shorter fs pulse.

In Fig. ~1 we depict the physical processes in the rescattering
\begin{figure}
\includegraphics[width=\figwidth]{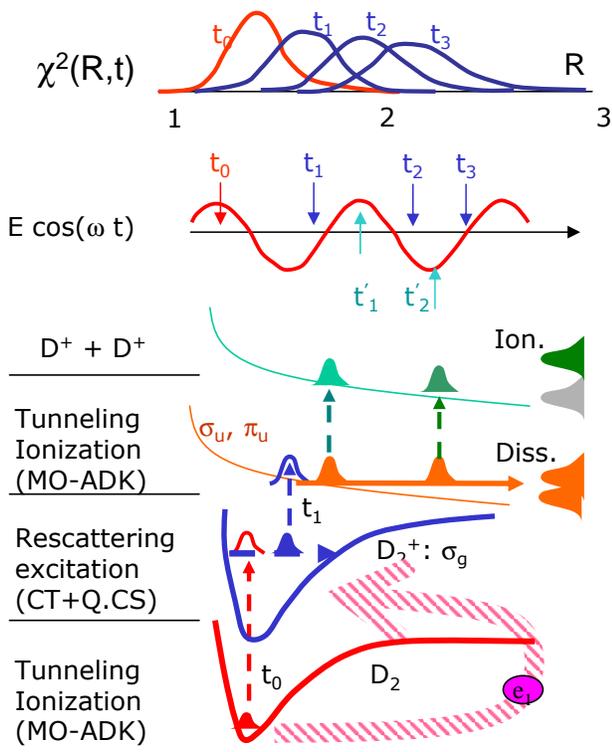}
\caption{\label{fig:1}  Schematic of the major physical processes
leading to the formation of D$^+$ ion. The D$_2$ is first ionized
at t$_0$ creating an electron wavepacket which returns to collide
with D$_2^+$ at time t$_1$. In the meanwhile the initial
vibrational wave packet, measured by $\chi^2$(R,t), created at
t$_0$, is shifted to larger R and broadened at later time. At
t$_1$, the D$_2^+$ is excited from $\sigma_g$ to $\sigma_u$ and
$\pi_u$ by electron impact. The excited D$_2^+$ can dissociate
directly to give D$^+$, or can be further ionized at t'$_1$,
t'$_2$,  etc to produce two D$^+$ ions by Coulomb explosion. Note
that similar rescattering processes can be initiated at later
time, t$_2$, t$_3$, etc, and are included in the calculated D$^+$
spectra. }
\end{figure}
mechanism which are used for reading a ``molecular clock''.
Similar to the experimental arrangement of I and II, the D$_2$
molecules are aligned perpendicular to the direction of a linearly
polarized infrared laser. This configuration eliminates D$^+$ ions
from being produced either through bond softening or through
charge resonance enhancement ionization \cite{Zou95a,Bandrauk99},
even though both processes produce D$^+$ ions at lower kinetic
energies which can be separated from the higher energy  ions
produced through the rescattering process \cite{Alnaser03}
considered here. Following Fig.~1, a D$_2$ molecule is ionized
near the peak field of the laser pulse at t$_0$. This ionization
launches two correlated wave packets: an electronic wave packet
which is driven by the combined laser field and the Coulomb field
of the D$_2^+$ ion, and a vibrational nuclear wave packet which is
assumed to propagate freely in the $\sigma_g$ ground potential of
D$_2^+$. The vibrational wave packet at t$_0$ is taken to be the
ground vibrational state of D$_2$, assuming Frank-Condon
principle. For $t > t_0$, the vibrational wave packet propagates
outward  and broadens, as shown in the top frame of Fig.~1. The
electron initially is driven out by the laser field but returns
after $2/3$ of  an optical cycle to recollide with D$_2^+$ at time
t$_1$. The rescattering  can excite D$_2^+$ to higher electronic
states or to ionize it. To  simplify the operation of the clock,
the laser intensity is  chosen such that the returning electron
has small energy and ionization is negligible. Once the D$_2^+$ is
in the excited electronic state, it can dissociate or it can  be
further ionized by the laser field. Since the laser pulse lasts
for many optical cycles, the electron can revisit the D$_2^+$ ion
many cycles later after the initial ionization. Thus rescattering
can occur on the second return at t$_2$, on the third return at
t$_3$, etc, see Fig.~1. For each initial t$_0$, the return times
$t_1, t_2, t_3$, ... are fairly well defined. Since the initial
ionization at t$_0$  occurs only in a sub-fs time interval,  the
rescattering times t$_i$ (i=1,2,3,...) are also defined at the
sub-fs (or attosecond) accuracy. If the rescattering populates
only the lowest excited electronic state, $\sigma_u$, and then the
D$_2^{+}$ dissociates, the peaks of the kinetic energy released
can be used to read the clock at the rescattering times, as
depicted in Fig.~1.

We have examined the rescattering model in details and found three
major modifications to the simple model used in I and II. First,
we found that excitation to the excited  $\pi_u$ electronic state
is not negligible in general. Second, once the D$_2^+$ ions are in
the excited electronic states, they are readily ionized by the
laser fields such that the D$^+$ ions are predominantly produced
from the  Coulomb repulsion. This modifies the D$^+$ ion energy
distributions. Third, we found that rescattering from the third
return is more important than from the first return, in
disagreement with  the model of I and II. However, our conclusion
is identical to the rescattering model results of He where
multiple returns have been shown to be dominant \cite{Yudin01}.

In the modelling, we calculate the ionization rate at t$_0$ using
the molecular tunneling (MO-ADK) model for a D$_2$ molecule
aligned perpendicular to the laser polarization \cite{Tong02b}.
The subsequent trajectory of the ionized electron is calculated
classically in the field of D$_2^+$ and the laser, with initial
longitudinal and transverse velocity distributions determined
according to the ADK theory. The return energy and return time of
the electron at the distance of closest approach are calculated.
To obtain electron impact excitation probability from $\sigma_g$
to the excited $\sigma_u$ and $\pi_u$ electronic states of
D$_2^+$, we used scaling relation similar to that used for He
\cite{Yudin01}, taking the 1s to 2p excitation cross section for
electron impact on H and He$^+$ as input data \cite{Bray}. Since
D$_2^+$ is aligned perpendicular to the electron beam direction,
we use the 2p$_1$ excitation cross section for the $\sigma_u$
state and the 2p$_0$ excitation cross section for the $\pi_u$
state, but each rescaled with their respective excitation
energies. It is interesting to note that electron impact
excitation to 2p$_0$ is five to three times larger than excitation
to 2p$_1$ in the energy region of interest (the branching ratio
for data from Bray \cite{Bray03} for H and from the experimental
data of \cite{Merabet99} for He are essentially identical). Thus,
different from I and II, we found that excitation to $\pi_u$ state
is not negligible. Once the D$_2^+$ ions are in the excited
electronic states, they can dissociate or further ionized by the
laser. We use the MO-ADK model \cite{Tong02b} to calculate the
ionization rate of D$_2^+$ in the $\sigma_u$ and $\pi_u$ states,
at different internuclear separations and for molecules aligned
perpendicular to the laser polarization. Since the electron cloud
of the $\pi_u$ state for the perpendicularly aligned D$_2^+$ is
along the laser field direction, it is readily ionized. For the
$\sigma_u$ the ionization rates are smaller but increase rapidly
with laser intensity. Using the ionization rates which are large
only near $t^{'}_1, t^{'}_2$, ... (see Fig.~1) following the
excitation of D$_2^+$ at  $t_1, t_2$, etc, we calculated the
ionization and dissociation yield and the D$^+$ kinetic energy
distributions. Since the vibrational wave packet spreads as time
increases, we need to weight the R-dependence  at each time in
calculating the ionization and dissociation yields, as well as the
corresponding energy of D$^+$. In the simulation, we also have to
integrate t$_0$ near the peak of the field where ionization of
D$_2$ was initiated.

\begin{figure}
\includegraphics[width=\figwidth]{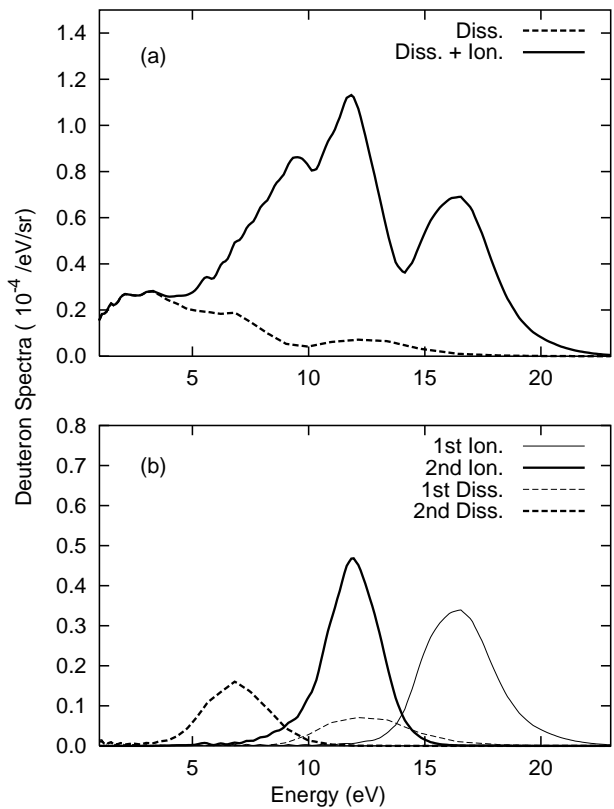}
\caption{\label{fig:2}  (a) The D$^+$ kinetic energy spectra
predicted by the present rescattering model. The dashed lines give
contribution from dissociation alone.   (b) Solid lines: D$^+$
spectra from the ionization channel, from the rescattering in the
first (higher energy peak) and 2nd optical cycles (lower energy
peak) after the initial ionization of D$_2$. Dashed lines, the
same except from the dissociation. Laser parameters used: Peak
intensity is $1.5\times 10^{14}$ W/cm$^2$, pulse length is 40 fs
and mean wavelength is 800 nm. }
\end{figure}

Figure~\ref{fig:2}(a) shows the total D$^+$ spectra obtained from
the present rescattering model. The total yield from ionization
and dissociation is shown in solid line. Note that the
dissociation alone, shown in dashed lines, contributes little to
the ion yield, except at lower energies. In Fig.~\ref{fig:2}(b) we
identify the dissociation and ionization yields from rescattering
at different times. This is important since it determines how
precisely a molecular clock can be read. In the figure we separate
the contributions into 1st and 2nd cycles. Each cycle is defined
to be one full optical cycle after the electron is born.   Thus
the ionization peak from the first cycle measures D$^+$ ion from
rescattering occurring near t$_1$, while the ionization yield from
the second cycle is due to rescattering occurring mostly at t$_2$
and t$_3$. Since the return energy at t$_2$ is smaller,
contribution in the second cycle comes mostly from rescattering at
t$_3$. From Fig.~\ref{fig:2}(b),   the peak position for the 3rd
return  in the ionization spectra essentially coincides with the
peak position from the first return in the dissociation spectra.
This peak in the total D$^+$ ion spectra was identified in II as a
reading of the clock at t$_1$, while our analysis shows that this
should be a reading of  t$_3$. (Note that the spectra in
Fig.~\ref{fig:2}(a) include contributions from higher returns that
are not shown in Fig.~\ref{fig:2}(b).)

In Fig.~3(a) we show the calculated D$^+$ spectra at four
different
\begin{figure}[b]
\includegraphics[width=\figwidth]{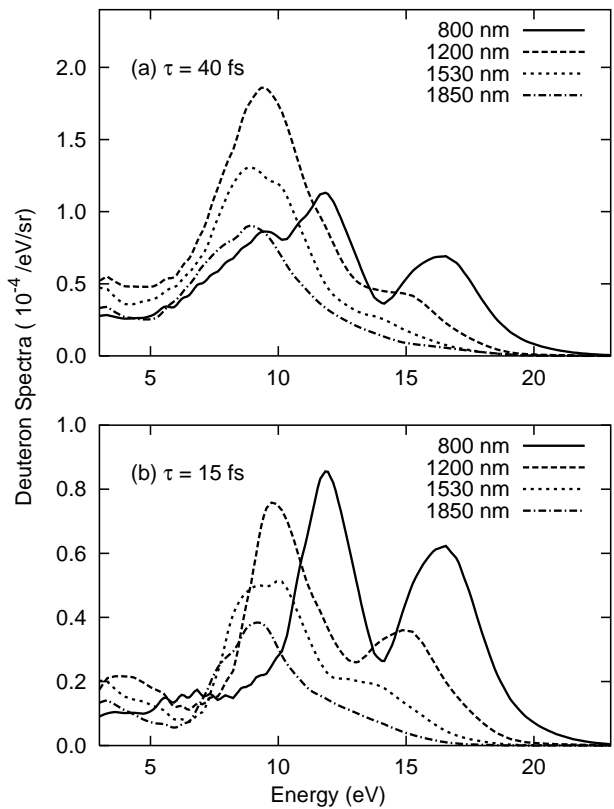}
\caption{\label{fig:3} (a) The dependence of the D$^+$ kinetic
energy spectra on the mean laser wavelength. The peak from the
first return is clearly visible only for the 800nm pulse. Laser
peak intensity is $1.5\times 10^{14}$ W/cm$^2$ and pulse length is
40 fs. (b) Same as (a) except that the pulse length is reduced to
15 fs. }
\end{figure}
wavelengths used in II. The data are to be compared to their
Fig.~2 except that we plot against the sum energy which is twice
the energy per D$^+$ used in II. The comparison shows reasonable
agreement, but not in details, since the modelling does not
account for the volume effect and the finite acceptance angle used
in the experiment. However, the shift of the main peak with
increasing wavelength is clear. In Table~\ref{tab:1} we compare
the data
\begin{table}
\caption{\label{tab:1} Comparison of the D$^+$ kinetic energy peak
positions (E in eV), and the deduced internuclear separation (R,
in a.u.) and the time (t, in fs) where rescattering occurs. The
experimental data and model results quoted in II \cite{Niikura03}
are shown together with the predictions from the present work.}
\begin{ruledtabular}
\begin{tabular}{||r||r||r|r|r||r|r|r||}
       & Exp. (II) & \multicolumn{3}{c||}{Model (II)} & \multicolumn{3}{c||}{This Work} \\
\hline $\lambda$ (nm)& E & E & R & t$_1$ & E & R & t$_3$ \\
\hline
 800  & 12.0  & 12.6 &1.68& 1.7 &  12.0 & 2.27 & 4.4 \\
 1200 & 10.0  & 10.7 &1.84& 2.6 &   9.4 & 2.90 & 6.7 \\
 1530 & 8.6   & 9.7  &1.94& 3.3 &   8.7 & 3.12 & 8.6 \\
 1850 & 8.0   & 8.1  &2.10& 4.4 &   8.2 & 3.30 & 10.5 \\
\end{tabular}
\end{ruledtabular}
\end{table}
deduced from II with the data deduced from our calculations shown
in Fig.~3(a). The experimental peak energy for each wavelength is
shown in column 2. Assuming that the peak is due to dissociation
for rescattering at t$_1$, the expected kinetic energy, the mean
internuclear separation and the rescattering time t$_1$ were
derived in II, and are shown in the third column. According to our
model, the peak at each wavelength is due to rescattering at
t$_3$, obtained from the ionization signal. The predicted deuteron
energies from our modelling , shown in the 4th column, are in good
agreement with the experimental data. However, we read a different
mean internuclear separation and a different rescattering time.
From this Table, we note that these measurements imply a
sub-angstrom resolution in the internuclear separation or an
attosecond resolution in the recollision time.

In Fig.~3(a) the peak positions for the 1530 nm and 1850 nm data
points are not very sharp. From Table I we note that t$_3$ for
each case is 8.6 fs and 10.5 fs respectively. At such long time
after the wave packet is created, the spreading of the vibrational
wave packet is large such that the clock can no longer be well
read. In fact, the peak energy and the mean internuclear
separation for these two wavelengths are calculated from the third
return alone, not from the full spectra. Thus the longer
wavelength laser blurs the reading of the clock.

Another issue that affects the reading of the clock is the pulse
duration.  A clock designed to read t$_1$ appears preferable since
the vibrational wave packet is not significantly broadened within
one optical cycle. However, at t$_1$ the mean internuclear
separation is small and the molecular potential curves are steep.
Upon reflection, this results in a rather broad D$^+$ kinetic
energy spectrum, see the dissociation or ionization peaks from the
first return in Fig.~2(b). Signals from the first return are also
weak, and the structures are easily buried by contributions from
the third return.  Indeed, our model indicates that the  kinetic
energy spectra are dominated by rescattering from the third
return. Even though the vibrational wave packet is broadened more
at t$_3$,  the breakup occurs at larger internuclear separations
where the potential curves are rather flat, resulting  in sharper
deuteron energy spectra, see Fig.~2(b). Unfortunately for a 40 fs
pulse used in I and II, higher returns ( for $t > t_3$) still
contribute to the rescattering process. These higher returns
modify the ion yield in the lower energy region, thus shifting the
overall peak position. By using a shorter laser pulse,
contributions from these higher multiple returns can be
eliminated. In Fig.~3(b) we show the calculated ion yield for a 15
fs laser pulse at the same intensity as in Fig.~3(a). Clearly the
main peak in the ion yield is more prominent and thus the time
resolution is improved. Note that the clock is not expected to be
affected by the lack of knowledge of the absolute phase of the
few-cycle laser pulse since the clock is measured from the time
t$_0$ where D$_2$ is first ionized. However, our analysis does not
include volume effect, nor the acceptance angle of D$^+$ ions,
which can reduce the precision of the molecular clock.

In summary, we examined the rescattering mechanism for the
production of high energy D$^+$ ions in laser-D$_2$ interaction.
We show that the D$^+$ ions are produced mainly not from the
dissociation of D$_2^+$ after it is excited by the returning
electron, but rather by  Coulomb explosion after the excited
D$_2^+$ is further ionized by the laser. We agree with Niikura
{\it et al.\ }\cite{Niikura02, Niikura03} that the ion spectra can
be used as a molecular clock with attosecond precision, and
suggest that experiments with a shorter fs (about 15fs) laser
pulse would improve the precision of the clock.

This work is in part supported by Chemical Sciences, Geosciences
and Biosciences Division, Office of Basic Energy Sciences, Office
of Science, U. S. Department of Energy. CDL also wishes to thank
Igor Bray for communicating to him the partial 1s
$\rightarrow$2p$_m$ (m=0,1) cross sections.


\end{document}